\documentclass[12pt]{article}

\usepackage{amsfonts}
\usepackage{epsfig}
\usepackage{latexsym}
\usepackage{graphicx}
\def\bequ{\begin{equation}}
\def\eequ{\end{equation}}
\def\barr{\begin{array}}
\def\earr{\end{array}}
\def\half{{1\over 2}}
\def\ben{\begin{equation}}
\def\een{\end{equation}}
\def\bena{\begin{eqnarray}}
\def\eena{\end{eqnarray}}

\renewcommand{\theequation}{\arabic{section}.\arabic{equation}}
\def\spa#1{\phantom{\fbox{\rule[-#1cm]{0cm}{0cm}}}}

\def\drawbox#1#2{\hrule height#2pt
        \hbox{\vrule width#2pt height#1pt \kern#1pt
              \vrule width#2pt}
              \hrule height#2pt}

\def\Fund#1#2{\vcenter{\vbox{\drawbox{#1}{#2}}}}
\def\Asym#1#2{\vcenter{\vbox{\drawbox{#1}{#2}
              \kern-#2pt       % line up boxes
              \drawbox{#1}{#2}}}}

\def\funda{\Fund{6.5}{0.4}}

\def\symm{\funda\kern-0.4pt\funda}

\def\b1{e^0}

\newcommand{\be}{\begin{equation}}
\newcommand{\ee}{\end{equation}}
\def\bea{\begin{eqnarray}}
\def\eea{\end{eqnarray}}
\def\Tr{\mbox{Tr}}

\def\del {\partial}
\def\nn{\nonumber}

\def\half {{1 \over 2}}
\def\be{\begin{equation}}
\def\ee{\end{equation}}
\def\bea{\begin{eqnarray}}
\def\eea{\end{eqnarray}}

\def\lesssim{\mathrel{\hbox{\rlap{\hbox{\lower4pt\hbox{$\sim$}}}\hbox{$<$}}}}
\def\gtrsim{\mathrel{\hbox{\rlap{\hbox{\lower4pt\hbox{$\sim$}}}\hbox{$>$}}}}

\renewcommand{\thefootnote}{\fnsymbol{footnote}}
\begin{document}

\hfuzz=100pt
\title{{\Large \bf{Anomalous radius shift in AdS$_4$/CFT$_3$}}}
\date{}
\author{Oren Bergman$^a$\footnote{bergman@physics.technion.ac.il} and Shinji Hirano$^b$\footnote{hirano@nbi.dk}
  \spa{0.5} \\
$^a${{\it Department of Physics, Technion}},
\\ {{\it Israel Institute of Technology}},
\\ {{\it Haifa 32000, Israel}}
  \spa{0.5} \\
$^b${{\it The Niels Bohr Institute}}
\\ {{\it Blegdamsvej 17, DK-2100 Copenhagen}}
\\ {{\it Denmark}}
}
\date{}

\maketitle
\centerline{}

\begin{abstract}
We study higher order corrections to the radius/M2-brane charge of $AdS_4\times S^7/\mathbb{Z}_k$. There are two sources of corrections: one from the orbifold singularity 
of $\mathbb{C}^4/\mathbb{Z}_k$, and the 
other from the discrete torsion associated with the homology 3-cycle 
$H_3(S^7/\mathbb{Z}_k,\mathbb{Z})=\mathbb{Z}_k$. 
We give a precise formula for the charge shift. These corrections are relevant, 
for example, at two loops in the $AdS_4\times\mathbb{C}P^3$ sigma model, 
and therefore for the strong coupling test of the all loop Bethe ansatz.   
\end{abstract}

\renewcommand{\thefootnote}{\arabic{footnote}}
\setcounter{footnote}{0}

%%%%%%%%%%%%%%%%%%%%%%%%%%%%%%%%%%

\newpage

\section{Introduction and summary}

It has recently been proposed that the superconformal field theory describing 
$N$ coincident M2-branes at the fixed point of the orbifold $\mathbb{C}^4/\mathbb{Z}_k$
(including $k=1$, which is flat space) is an ${\cal N}=6$ supersymmetric Chern-Simons-matter
theory with a gauge group $U(N)_k\times U(N)_{-k}$, and matter in the
bi-fundamental representation \cite{Aharony:2008ug}.
At large $N$ and $k$ this provides explicit realizations of $AdS_4/CFT_3$ duality
\cite{Maldacena:1997re}. This class of AdS/CFT dualities is somewhat different
from the $AdS_5/CFT_4$ case, since the orbifold plays an essential role in its
formulation. The orbifold provides an integer parameter $k$,
which corresponds in the dual CFT to the level of the CS terms.
On the other hand the three-dimensional CFT does not have any continuious
parameter like the four-dimensional $g_{YM}$.
$AdS_5/CFT_4$ duality can be extended to orbifolds, but these 
preserve at most half the supersymmetry, and generically break all of it.
In the $AdS_4/CFT_3$ case the CFT action has the same amount of supersymmetry 
for all $k$.\footnote{For $k=1,2$ the supersymmetry is enhanced non-perturbatively
to ${\cal N}=8$ \cite{Aharony:2008ug}.}

The parameter $k$ also gives $AdS_4/CFT_3$ a
somewhat richer structure than $AdS_5/CFT_4$. 
First, there are actually two supergravity duals that have different regimes
of validity, which depend on the relative scaling of $k$ and $N$.
For $k\ll N^{1/5}\ll N$ the supergravity dual can be described as M theory 
on $AdS_4\times S^7/\mathbb{Z}_k$, whereas for $N^{1/5} \ll k \ll N$
the appropriate description is in terms of Type IIA string theory on
$AdS_4\times \mathbb{C}P^3$.
In the latter one would naively expect the dilaton to provide a continuous parameter in the
field theory (as in the four-dimensional case), however the dilaton is fixed
by $N$ and $k$ in this background.
Second, the structure of the internal space is richer, and this allows, for example, 
the possibility of turning on a discrete flux, which changes the relative ranks of the two gauge group factors \cite{Aharony:2008gk}.

In this paper we will exhibit another interesting difference between the the three
and four-dimensional versions of $AdS/CFT$.
The maximally supersymmetric $AdS_5\times S^5$ background in Type IIB 
string theory is believed to be exact,
namely free of any higher order corrections \cite{Metsaev:1998it}.
This was also argued to be true for the maximally supersymmetric $AdS_4\times S^7$ 
(and $AdS_7\times S^4$) background in M theory \cite{Kallosh:1998qs}.
However we will show that for $AdS_4\times S^7/\mathbb{Z}_k$ with $k>1$ 
there is a higher order (in the curvature) correction to the 
background.\footnote{The possibility of corrections were mentioned in footnote 8 of \cite{McLoughlin:2008he}.}
An important consequence of this correction is that the radius of curvature in 
the Type IIA description is shifted to
\be
\label{main_result}
R_{str}^2=2^{5/2}\pi\sqrt{\lambda -{1\over 24}\left(1-{1\over k^2}\right)+{l^2\over 2k^2}} \ ,
\ee
where $0\leq l \leq k-1$ is the number of discrete torsion flux quanta.
In particular, this indicates that the shift becomes relevant at two loops  ${\cal O}(1/\sqrt{\lambda})$ in the $AdS_4\times \mathbb{C}P^3$ sigma model. Thus it affects, for example, two loop corrections to the energy/anomalous dimension of giant magnons and spinning strings. 
Hence the radius shift is important for the strong coupling test of the all loop Bethe ansatz proposed in \cite{Gromov:2008qe} (see also \cite{Minahan:2008hf,Bak:2008vd,Zwiebel:2009vb} for a list of papers
on integrability in this model).

The source of this correction is 
the orbifold singularity in the original background $\mathbb{C}^4/\mathbb{Z}_k$.
It is well known that orbifold singularities in M theory can be a source of M2-brane charge
via the gravitational coupling $\int C_3\wedge I_8$, where $I_8$ is a curvature
8-form \cite{Sen:1996zq, Dasgupta:1997cd, Sethi:1998zk}.
We will compute this charge for $\mathbb{C}^4/\mathbb{Z}_k$, and an additional 
correction due to discrete torsion. 
This will generalize the result of \cite{Sethi:1998zk}, where the charge was computed
for the $\mathbb{Z}_2$ case.
The shift in the M2-brane charge leads in the large $N$ limit to the shift
in the radius of the near-horizon geometry (\ref{main_result}).
These corrections are subleading in the supergravity approximation,
and were not included in \cite{Aharony:2008ug, Aharony:2008gk}.

In section 2 we will review the relevant features of the ABJM model and its supergravity duals.
In section 3 we will derive our main result (\ref{main_result}) by computing the M2-brane
charge shift due to the orbifold singularity and to discrete torsion.
We will also interpret these corrections in the Type IIA description in section 4.

\medskip

\noindent\underline{Note added}: There may be a correction to the radius shift in (\ref{main_result})
where $l \rightarrow l - {k\over 2}$, due to a possible parity anomaly for M2-branes in this background.
This question will be explored elsewhere \cite{forthcoming}.

\setcounter{equation}{0}
\section{Essential features of the ABJM model}

We will be concerned with the simplest case of the $AdS_4/CFT_3$ duality, namely
the one corresponding to $N$ M2-branes placed at a $\mathbb{C}^4/\mathbb{Z}_k$
orbifold singularity. As shown in \cite{Aharony:2008ug},
the three-dimensional worldvolume field theory on the 
M2-branes is an ${\cal N}=6$ supersymmetric Chern-Simons theory,
that has a gauge group $U(N)\times U(N)$, with CS levels $(k,-k)$,
as well as matter superfields transforming in the $({\bf N},\bar{\bf N})$ 
representation of the gauge group and in the ${\bf 4}$ of the $SU(4)_R$
R-symmetry group.
This is an interacting superconformal field theory with a coupling constant 
given by $1/k$. At large $N$ and $k$ the theory has an 'tHooft (planar) limit with
fixed $\lambda \equiv N/k$, and the field theory description is (perturbatively)
valid when $\lambda \ll 1$, {\em i.e.} for $k\gg N$.

For $k\ll N$ there is a dual supergravity description 
given by the near horizon limit of the M2-branes on $\mathbb{C}^4/\mathbb{Z}_k$.
The background corresponding to $N$ M2-branes on $\mathbb{C}^4/\mathbb{Z}_k$
has a metric and 4-form field strength given by
\bea
ds_{11}^2&=&H(z_I)^{-2/3}(-dt^2+dx_1^2+dx_2^2)
+H(z_I)^{1/3}ds^2_{\mathbb{C}^4/\mathbb{Z}_k} \nonumber\\
G_4&=&dt\wedge dx_1\wedge dx_2\wedge dH^{-1} \ , 
\label{11d_background}
\eea
where $z_I\in \mathbb{C}^4$ ($I=1,\ldots,4$), and $H$ is the harmonic function on 
$\mathbb{C}^4$,
\be
H(z^I) = 1 + {Q\over r^6} \ , 
\ee
with $r^2=z^I z^I$, and $Q=32\pi^2(kN)\ell_p^6$.
In the near horizon limit this becomes $AdS_4\times S^7/\mathbb{Z}_k$,
\bea
ds_{11}^2 &=& {R^2\over 4}ds^2_{AdS_4}+R^2ds^2_{S^7/\mathbb{Z}_k} \nonumber \\
G_4 &=& {3\over 8}R^3 \epsilon_4\ ,
\label{11d_near_horizon}
\eea
where $R=(2^5 \pi^2 kN)^{1/6}\ell_p$, and $\epsilon_4$ is the unit volume form on $AdS_4$.
The metric on $S^7/\mathbb{Z}_k$ is conveniently described in terms of the Hopf
fibration over $\mathbb{C}P^3$ as
\be
\label{Hopf}
ds^2_{S^7/\mathbb{Z}_k} = {1\over k^2}(d\varphi + k\omega)^2 + ds^2_{\mathbb{C}P^3} \ ,
\ee
where $ds^2_{\mathbb{C}P^3}$ is the Fubini-Study metric on $\mathbb{C}P^3$,
$\varphi$ is a periodic coordinate with period $2\pi$, and $\omega$ is a one-form
related to the Kahler form on $\mathbb{C}P^3$ by $d\omega = J$.
Since the radius of the circle is given by $R/k \sim (N/k^5)^{1/6}$, 
the eleven-dimensional supergravity description is only valid in the range $k\ll N^{1/5}$.

In the range $N^{1/5} \ll k \ll N$ one should really use the ten-dimensional 
Type IIA supergravity description, which is given by the dimensional reduction of 
(\ref{11d_near_horizon}) (setting $\alpha'=1$) \cite{Nilsson:1984bj,Watamura:1983hj} :
\bea
ds^2_{10} &=& {R^3\over k }\left({1\over 4}ds^2_{AdS_4} + ds^2_{\mathbb{C}P^3}\right) 
\ , \  e^{2\Phi} = {R^3\over k^3} \ , \ F_2 = kJ \nonumber \\
\tilde{F}_4 &=& {3\over 8} R^3 \epsilon_4 \ .
\label{10d_background}
\eea
The radius of curvature in string units is $R_s^2 = R^3/k = 2^{5/2} \pi (N/k)^{1/2}$,
and the string coupling is $g_s=e^{\Phi} = 2^{5/4} \pi^{1/2} (N/k^5)^{1/4}$,
so we see that the Type IIA supergravity description is indeed valid
for $N^{1/5} \ll k \ll N$.

The simplest generalization of this story is to change the gauge group to
$U(M)\times U(N)$ with $M\neq N$ \cite{Hosomichi:2008jb,Aharony:2008gk}. 
This was shown to correspond to turning on
discrete torsion in the orbifold $\mathbb{C}^4/\mathbb{Z}_k$ \cite{Aharony:2008gk}.
Since this takes values in $H_3(S^7/\mathbb{Z}_k,\mathbb{Z}) = \mathbb{Z}_k$, it suggests
that there are only $k$ distinct ${\cal N}=6$ superconformal CS theories with minimal rank $N$.
Indeed evidence was presented in \cite{Aharony:2008gk} that there are no superconformal 
theories with $|M-N|>k$, and that the theories with $M-N \leq 0$ are related to those with 
$M-N \leq k$
by a three-dimensional version of Seiberg duality (called parity duality).
The distinct theories are therefore $U(N+l)_k \times U(N)_{-k}$, with $l=0,\ldots, k-1$.
At large $N$ and $k\ll N^{1/5}$ these theories are dual to the M theory background 
(\ref{11d_near_horizon}), with an additional discrete holonomy for the $C$ field,
\be
\label{C_holonomy}
\int_{S^3/\mathbb{Z}_k\subset S^7/\mathbb{Z}_k} {C_3\over 2\pi} =  {l\over k}\ .
\ee
In the Type IIA description, which is valid when $N^{1/5}\ll k \ll N$, this becomes a $B$ field
holonomy on $\mathbb{C}P^1 \subset \mathbb{C}P^3$,
\be
\label{B_holonomy}
\int_{\mathbb{C}P^1\subset\mathbb{C}P^3} {B_2\over (2\pi)^2} = {l\over k} \,.
\ee
As explained in \cite{Aharony:2008gk}, although the $B$ field holonomy is not quantized topologically
(unlike the $C$ field holonomy in the M theory description), it is quantized dynamically.

\setcounter{equation}{0}
\section{Anomalous M2-brane charges}

We will now show that the solution describing $N$ M2-branes on
$\mathbb{C}^4/\mathbb{Z}_k$ (\ref{11d_background}) receives a correction
due to the fact that the fixed plane itself carries a certain M2-brane charge.
The bosonic part of the M theory low energy effective action is given by
\be
S_{11}={1\over 2\kappa_{11}^2}\left[\int d^{11}x\,\sqrt{-G}\left(R-\half |G_4|^2\right)
-{1\over 6}\int C_3\wedge G_4\wedge G_4+(2\pi)^4\beta\int C_3\wedge I_8\right]\,,
\ee
where $\beta$ is related to the 5-brane tension by $T_6=1/(2\pi)^3\beta$ and $I_8$ is an 8-form anomaly polynomial \cite{Duff:1995wd}, which is given in terms of Pontryagin classes as \cite{AlvarezGaume:1983ig}
\be
I_8=-{1\over 2\cdot 4!}\left(p_2 - {1\over 4}p_1^2\right)\ ,
\ee
with 
\bea
p_1=- {1\over 2(2\pi)^2}\Tr R^2\quad\mbox{and}\quad
p_2= {1\over 8(2\pi)^4}\left[ (\Tr R^2)^2 - 2\Tr R^4 \right] \,.\nn
\eea
For a compact manifold the 8-form $I_8$ is related to the Euler class by
\be
\int_{{\cal M}_8}I_8 =-{\chi\over 24}\ . 
\ee
For non-compact manifolds the Euler class has a boundary contribution,
and $I_8$ is related only to the bulk part.
We can see from the $C_3$ equation of motion that there are three possible types of 
contributions to the M2-brane charge, 
\be
d\star G_4=(2\pi)^2N\delta^8(x)-\half G_4\wedge G_4+(2\pi)^2I_8\ ,\label{charges}
\ee
where we have set $\beta=1/(2\pi)^2$.
The total charge is given by the integral over the 8-manifold
\be
Q_{M2} = {1\over (2\pi)^2}\int_{\del{\cal M}_8}\star\, G_4 
=N -{1\over 2(2\pi)^2}\int_{{\cal M}_8}G_4\wedge G_4 -{\chi_{bulk}\over 24}\ .
\ee
The three terms correspond respectively to the contributions of M2-brane sources, flux,
and the (bulk) geometry of the 8-manifold.\footnote{For a compact 8-manifold this has 
to vanish, 
and this leads to the anomaly cancellation condition
(the M-theory analog of tadpole cancellation) \cite{Sethi:1996es,Witten:1996md},
\bea
N-{1\over 2(2\pi)^2}\int_{{\cal M}_8}G_4\wedge G_4-{\chi\over 24}=0\ .\nn
\eea
}

\subsection{The orbifold contribution}

Let us begin with the contribution of the geometry, $\mathbb{C}^4/\mathbb{Z}_k$. 
The total Euler number of this space is given by \cite{Mohri:1997ef}
\be
\chi(\mathbb{C}^4/\mathbb{Z}_k)= k\ .
\ee
This has a contribution from the bulk due to the fixed point,
and a contribution from the boundary $S^7/\mathbb{Z}_k$.
The contribution of the boundary is easily computed by realizing that $\mathbb{Z}_k$ acts freely
on $S^7$, which is the boundary of $\mathbb{C}^4$, and by the fact that
$\chi_{bnd}(\mathbb{C}^4) = \chi(\mathbb{C}^4) = \chi(point) = 1$.
Therefore $\chi_{bnd}(\mathbb{C}^4/\mathbb{Z}_k)=1/k$, and the contribution 
of the fixed point can be easily extracted:
\be
\chi_{bulk}(\mathbb{C}^4/\mathbb{Z}_k)=\chi(\mathbb{C}^4/\mathbb{Z}_k) - \chi_{bnd}(\mathbb{C}^4/\mathbb{Z}_k)
= k - {1\over k}\ .\label{conjecture}
\ee
Alternatively, the contribution of the fixed point can be computed directly using 
string theory by replacing $\mathbb{C}^4/\mathbb{Z}_k$ with $T^8/\mathbb{Z}_k$.
Of course this is only possible for $k=2,3,4$ and $6$, since $\mathbb{Z}_k$ must be an automorphism
of the lattice defining the $T^8$.
The Euler numbers of $T^8/\mathbb{Z}_k$ were computed in \cite{Font:2004et}. 
In Appendix A we extract from these the Euler numbers of the fixed points
for these four cases, and show that they agree with the general result 
above.\footnote{The same formula holds in four dimensions for the ALE space
$A_{k-1}$, which has an orbifold limit $\mathbb{C}^2/\mathbb{Z}_k$.
For example the Eguchi-Hanson space ($A_1$) has $\chi(EH) = 2$,
which is made up of a boundary ($S^3/Z_2$) contribution 
$\chi_{bnd}(EH)=1/2$, and a bulk contribution $\chi_{blk}(EH)=3/2$.
Alternatively, one can compute the bulk contribution by considering $T^4/Z_2$,
which is an orbifold limit of $K3$. From $\chi(K3)=24$, and the fact that
the compact orbifold has a total of $2^4=16$ fixed points (and no boundary),
we find that $\chi_{fp} = 24/16 = 3/2$, in agreement with the bulk contribution
in the EH space.}

We conclude that the fixed point of the orbifold $\mathbb{C}^4/\mathbb{Z}_k$ carries
an M2-brane charge:
\be
Q_{M2}(\mathbb{C}^4/\mathbb{Z}_k) = -  {1\over 24}\left(k-{1\over k}\right) \ .
\ee
This agrees with, and generalizes, the result of \cite{Sethi:1998zk} for the ``$OM2^-$-plane"
$\mathbb{R}^8/\mathbb{Z}_2$,
\be
Q_{M2}(OM2^-) = -{1\over 16} \ .
\ee
In this special case there is an additional consistency check which comes from
compactifying one of the coordinates of the $\mathbb{R}^8$,
and reducing to Type IIA string theory.
There are two $OM2^-$-planes in this case, that become a single orientifold plane
$O2^-$ in Type IIA string theory. The D2-brane charge of the orientifold plane can 
be computed
independently using string theory, and the result is $-1/8$, percisely twice
the charge of the $OM2^-$-plane.
There is no analogous simple Type IIA reduction for $k>2$.

\subsection{The discrete torsion contribution}

The contribution of the discrete torsion to the M2-brane charge comes from the flux term:
\be
Q_{M2}^{torsion}= - {1\over 2} \int_{\mathbb{C}^4/\mathbb{Z}_k} {G_4\over 2\pi} 
\wedge {G_4\over 2\pi} 
 = -\half\int_{S^7/\mathbb{Z}_k} {G_4\over 2\pi}\wedge {C_3\over 2\pi}\ ,
\ee
where the torsion class corresponds to the discrete holonomy of $C_3$ given in (\ref{C_holonomy}).
To evaluate this quantity we will generalize the approach of \cite{Sethi:1998zk},
where it was computed for the $\mathbb{Z}_2$ case.
We consider a smooth 8-manifold ${\cal M}$ whose boundary is $S^7/\mathbb{Z}_k$,
and express the charge as
\be
\label{dt_charge}
Q_{M2}^{torsion}=-\half\int_{{\cal M}} {G_4\over 2\pi}\wedge {G_4\over 2\pi}\ .
\ee
The holonomy of $C_3$ on the torsion 3-cycle $S^3/\mathbb{Z}_k$ can
likewise be expressed as
\be
\label{G_flux}
\int_{{\cal W}}{G_4\over 2\pi}={l\over k}\ ,
\ee
where ${\cal W}$ is a 4-dimensional submanifold of ${\cal M}$ whose 
boundary is  $S^3/\mathbb{Z}_k$.
A class $G_4$ in ${\cal M}$ that satisfies (\ref{G_flux}) can be constructed
from the Poincare dual of the base $\mathbb{C}P^3$,
which is a 2-form $X$ that satisfies
\be
\int_{{\cal W}} X\wedge X = -k\ . \label{XX}
\ee
We can therefore identify $G_4/(2\pi)=-(l/k^2)X\wedge X$,
and the discrete torsion contribution to the M2-brane charge is given by
\be
Q^{torsion}_{M2}=-{l^2\over 2k^4}\int_{{\cal M}}X\wedge X\wedge X\wedge X={l^2\over 2k}\ .
\label{M2chargedt}
\ee
This generalizes the result of \cite{Sethi:1998zk} for $k=2$.
In that case there is only one choice of discrete torsion $l=1$, corresponding to
an $OM2^+$-plane which carries an M2-brane charge of $3/16$.
Upon compactifying one of the $\mathbb{R}^8$ directions one then obtains one of the
four variants of orientifold 2-planes in Type IIA string theory, $O2^-$, $O2^+$,
$\widetilde{O2}^-$, or $\widetilde{O2}^+$, depending on which $OM2$-planes are placed
at opposite points on the circle. Again, a similar consistency check by reduction to ten
dimensions cannot be made for $k>2$.

\subsubsection{An explicit construction of ${\cal M}$ and $X$}

Let us now make this computation more explicit.
We take the metric on ${\cal M}$ to be\footnote{We parametrize the 
$\mathbb{C}P^3$ manifold as in Appendix B.}
\be
ds_{{\cal M}}^2= {dr^2\over 1-{1\over (2r)^k}} 
+ \left(1-{1\over (2r)^k}\right)(d\varphi+\omega)^2
+ ds_{\mathbb{C}P^3}^2 \ ,
\ee
where $\varphi\sim\varphi+2\pi/k$, and $r\geq 1/2$. 
This space is smooth at $r=1/2$, and its boundary at $r\rightarrow\infty$ is 
$S^7/\mathbb{Z}_k$. 
It is a fiber bundle over $\mathbb{C}P^3$, where the fiber is a disk $D^2$ endowed with 
the metric
\be
ds_{D^2}^2= {dr^2\over 1-{1\over (2r)^k}} 
+ \left(1-{1\over (2r)^k}\right)d\varphi^2\ .
\ee
The 2-form $X$ that we want has the general form
\be
X=f(r)dr\wedge (d\varphi+\omega)+g(r)J\ ,
\ee
where $f(\infty)=g(\infty)=0$.
Since we want to identify $G_4$ with $X\wedge X$ we require this 4-form
to be closed, which implies that\footnote{The 4-form is then locally exact
$X\wedge X =d\left(g^2 (d\varphi+\omega)\wedge J+ d\chi\right)$,
where $\chi$ is an a-priori arbitrary 2-form. 
Comparing the boundary value with the torsion 3-form in \cite{Aharony:2008gk}
shows that $\chi\propto d\varphi\wedge\omega$.}
\be
f(r)={dg(r)\over dr}\ .
\ee
Being the Poincare dual of the $\mathbb{C}P^3$ base implies that the integral
of $X$ on the disk is unity
\be
\int_{D^2}X  = {2\pi\over k}g(r)\Biggr|_{r=1/2}^{\infty}=1 \ ,
\ee
and therefore that 
\be 
g\left({1\over 2}\right) = - {k\over 2\pi} \ .
\ee
The 4-dimensional submanifold ${\cal W}$ is taken to be
\be
ds_{{\cal W}}^2= {dr^2\over 1-{1\over (2r)^k}} 
+{1\over 4} \left(1-{1\over (2r)^k}\right)(d\psi_1+\cos\theta_1d\phi_1)^2
 +{1\over 4}\left(d\theta_1^2+\sin^2\theta_1d\phi_1^2\right)\ .
\ee
The boundary of ${\cal W}$ is indeed the torsion 3-cycle $S^3/\mathbb{Z}_k$ in 
$S^7/\mathbb{Z}_k$ at $r\rightarrow\infty$, as defined in Appendix B. 
The integral of the 4-form $X\wedge X$ over ${\cal W}$ then gives
\be
\int_{{\cal W}}X\wedge X={(2\pi g)^2\over k}\Biggr|_{1/2}^{\infty}
= - k \ ,\label{XXW}
\ee
as stated in (\ref{XX}).
We have defined the orientation of $\mathbb{C}P^1$ with the metric 
$ds_{\mathbb{C}P^1}^2=d\theta_1^2+\sin^2\theta_1d\phi_1^2$ 
by $\int_{\mathbb{C}P^1}J>0$.
The integral of the 8-form $X^4$ over ${\cal M}$ is then straightforward 
to evaluate:
\be
\int_{{\cal M}}X\wedge X\wedge X\wedge X
={(4\pi g)^4\over 16k}\Biggr|_{1/2}^{\infty}=-k^3\ ,
\ee
where we have defined the orientation of $\mathbb{C}P^3$ by $\int_{\mathbb{C}P^3}J\wedge J\wedge J>0$.

\medskip

This concludes the computation of the discrete torsion contribution to the M2-brane charge (\ref{M2chargedt}). To summarize, the total M2-brane charge shift due to the orbifold 
and discrete torsion is therefore
\be
\label{shift}
\Delta Q_{M2} = -  {1\over 24}\left(k-{1\over k}\right) + {l^2\over 2k} \, ,
\ee
which leads to our main result (\ref{main_result}).

\setcounter{equation}{0}
\section{Type IIA interpretation}

In the previous section we computed the M2-brane charge shift from the M theory
orbifold geometry and discrete torsion.
In this section we would like to try to interpret this result from the point of view
of Type IIA string theory. 
As we mentioned above, in the $\mathbb{Z}_2$ case there is a very simple Type IIA
interpretation of the charges in terms of orientifold 2-planes.
However there is no analogously simple interpretation in the more general $\mathbb{Z}_k$ 
case.

It is instructive to first consider a simpler example. Suppose that the M2-branes are placed in the direct product of a Taub-NUT space and a 4-manifold, $M_8=TN\times M_4$. 
Reducing on the asymptotic $S^1$ in the TN space gives 
an {\it anti}-D6-brane wrapping 
$M_4$ in the Type IIA D2-brane background.\footnote{See Appendix C for our sign conventions.}
The integral of the 8-form $I_8$ for a product space simplifies to
\be
\int_{M_8}I_8 = {1\over 4\cdot 24}\,\int_{TN} p_1(TN) \int_{M_4} p_1(M_4)=-{1\over 48}\int_{M_4} p_1(M_4)\ ,
\ee
where we have used that the first Pontryagin number of the TN space is $-2$. 
This implies that the curvature CS coupling $C_3\wedge I_8$ localizes on the D6-brane worldvolume and reduces to a higher order gravitational CS coupling $+{1\over 48}C_3\wedge p_1(M_4)$ \cite{Sen:1997pr}.
Meanwhile, in the TN background the G-flux CS coupling $C_3\wedge G_4\wedge G_4$ also localizes on the D6-brane worldvolume and becomes the familiar 
CS coupling $C_3\wedge {\cal F}\wedge{\cal F}$, where 
${\cal F}=2\pi\alpha' F + B_2$  \cite{Imamura:1997ss}.
Hence in the presence of D6-branes the bulk CS couplings $C_3\wedge G_4\wedge G_4$ and $C_3\wedge I_8$  localize on the D6-brane worldvolume and 
reduce to (setting $\alpha'=1$)\footnote{The second term is 
a part of higher order curvature corrections expressed in terms of the $\hat{A}$-genus
\cite{Bershadsky:1995qy, Green:1996dd, Cheung:1997az}.}
\be
{\cal L}_{CS}= {1\over (2\pi)^6}\, C_3\wedge \left[\half {\cal F}\wedge{\cal F} 
-{1\over 48} {(2\pi)^4\over 8\pi^2}\Tr R\wedge R \right]_{M_4}\, . \label{D6CS}
\ee
Therefore both the worldvolume gauge field and the curvature of the 4-manifold
$M_4$ can induce D2-brane charge within the D6-brane.
In M theory this corresponds to a shift in the M2-brane charge.

In our case the 8-dimensional geometry does not have a simple product structure.
Before going to the 
(gauge theory) 
IR limit it corresponds to an intersection
of two different TN spaces (KK monopoles), 
or equivalently to a particular toric hyper-kahler space
$X_8$ \cite{Gauntlett:1997pk,Aharony:2008ug}. 
In the IR limit this reduces to $\mathbb{C}^4/\mathbb{Z}_k$.
The reduction to Type IIA string theory gives an intersection of a KK monopole
with a bound state of a KK monopole and $k$ D6-branes.
This suggests that one may be able to account for the M2-brane charge shift (\ref{shift}) from the D6-brane CS couplings (\ref{D6CS}). 
However the Type IIA background is really some 7-manifold with $k$ units
of RR 2-form flux, and does not have any actual D6-branes.
It therefore seems difficult to compute the charge shift using the Type IIA reduction
of the M theory background.

\subsection{D-brane domain wall probes}

Although it is difficult to compute the M2-brane charge shift (\ref{shift}) directly in the Type IIA
background, we can detect it using probe D-branes.
Let us consider the near-horizon Type IIA background $AdS_4\times \mathbb{C}P^3$ (\ref{10d_background}).
A D$p$-brane that wraps a cycle in $\mathbb{C}P^3$, and extends along 
all but the radial direction of $AdS_4$, forms a domain wall in $AdS_4$
across which some of the fluxes jump.
In particular a D4-brane which wraps $\mathbb{C}P^1$ creates a jump in 
$\hat{F}_4 = \tilde{F}_4 + B_2\wedge F_2$ (by changing $B_2$) and therefore in $l$, 
and a D6-brane which wraps $\mathbb{C}P^2$ creates a jump in $F_2$ and therefore in $k$.\footnote{A D8-brane
wrapping the whole $\mathbb{C}P^3$ would create a jump in $F_0$.
This is not part of our background, and is beyond the scope of our paper.}
In either case there is also a jump in $*\tilde{F}_4$ due to an induced D2-brane charge,
which should agree with the jump in the M2-brane charge shift from (\ref{shift}).
We do not expect to reproduce the M theory result precisely in this way, since the Type IIA
description is valid only for large $k$. But we do expect to get the leading order term in $1/k$ correct.

Let us start with the D4-brane.
A D4-brane wrapped on $\mathbb{C}P^1\subset \mathbb{C}P^3$, 
and localized at a fixed radial position $r_0$, forms a domain
wall in $AdS_4$ across which 
the flux of $\hat{F}_4$ on $\mathbb{C}P^2\subset \mathbb{C}P^3$ increases by one unit,
{\em i.e.} $l\rightarrow l+1$. 
The resulting jump in the M2-brane charge shift computed from (\ref{shift}) is given by
\be
\label{ljump}
\delta_l\Delta Q_{M2} = {(l + 1)^2\over 2k} - {l^2\over 2k} = {l\over k} + {1\over 2k} \,.
\ee
Note that since $l$ can be as large as $k-1$, this can include a leading order effect in $1/k$.
The D2-brane charge induced on the D4-brane is given by the $B_2$ field,
\be
\label{D2inD4}
Q_{D2}^{(D4)} = {1\over (2\pi)^2} \int_{\mathbb{C}P^1} B_2 =  {l\over k} \,,
\ee
which agrees with (\ref{ljump}) to leading order in $1/k$.\footnote{We used the value of the $B$ field
at $r<r_0$. The agreement is actually precise if we use instead the average value of the $B$ field
from the two sides. But this is irrelevant to leading order in $1/k$.}
%In fact we can reproduce the entire discrete torsion contribution to the M2-brane charge
%shift in this way, by inserting $l$ D4-brane probes in the ABJM background:
%\be
%Q_{D2}^{(D4)} = {l\over 2}\left({0 + l\over k} + 0 \right) = {l^2\over 2k} \,.
%\ee

For a D6-brane domain wall the flux of $F_2$ decreases by one unit, so $k\rightarrow k-1$.
The jump in the M2-brane charge shift is then 
\be
\label{kjump}
\delta_k\Delta Q_{M2} =  {1\over 24}\left(1 + {1\over k(k - 1)}\right) + {l^2\over 2k(k - 1)} \,,
\ee
where the first term comes from the contribution of the orbifold geometry, and the second
term comes from discrete torsion contribution.
Let us compare this with the D2-brane charge induced on the D6-brane.
There are two contributions.
The first is from the $B_2$ field,
\be
\label{D2inD6_B}
Q_{D2}^{(D6,B)} =  {1\over 2(2\pi)^4} \int_{\mathbb{C}P^2} B_2 \wedge B_2 =  {l^2\over 2k^2} \,,
\ee
which agrees to leading order in $1/k$ with the discrete torsion contribution in (\ref{kjump}).
The second contribution is from the curvature coupling (see Appendix C),
\be
Q_{D2}^{(D6,R)} = {1\over 48} \int_{\mathbb{C}P^2} \left(p_1(T(\mathbb{C}P^2)) 
- p_1(N(\mathbb{C}P^2))\right) \,,
\ee
where the two terms are the first Pontrjagin classes of the tangent and normal bundles, respectively,
of $\mathbb{C}P^2 \subset \mathbb{C}P^3$.
This can be computed as follows.
The total Pontrjagin class of the tangent bundle of $\mathbb{C}P^n$ is given by \cite{Eguchi:1980jx}
\be
p(T(\mathbb{C}P^n)) = (1+x^2)^{n+1} \,,
\ee
where $x$ is the generator of $H^2(\mathbb{C}P^n;\mathbb{Z})$,
which we can identify with the Kahler form $J$, if we assume the normalization $\int_{\mathbb{C}P^1} J = 1$.
In particular this gives
\be
p_1(T(\mathbb{C}P^n)) = (n+1) J\wedge J \,.
\ee
Now consider the submanifold $\mathbb{C}P^l \subset \mathbb{C}P^n$ ($l<n$).
Its tangent and normal bundles satisfy the Whitney sum relation,
$T(\mathbb{C}P^l) \oplus N(\mathbb{C}P^l) = T(\mathbb{C}P^n)$,
and therefore their Pontrjagin classes satisfy 
$p(T(\mathbb{C}P^l)) \wedge p(N(\mathbb{C}P^l)) = p(T(\mathbb{C}P^n))$.
In particular this implies 
\be
p_1(T(\mathbb{C}P^l)) + p_1(N(\mathbb{C}P^l)) = p_1(T(\mathbb{C}P^n)) \,,
\ee
and therefore that 
\be
p_1(N(\mathbb{C}P^l)) = (n-l) J\wedge J \,.
\ee
For our case this implies that
 \be
 Q_{D2}^{(D6,R)} = {1\over 48}(3-1) = {1\over 24} \,,
 \ee
 which agrees with the geometry contribution in (\ref{kjump}) to leading order in $1/k$.

\section*{Acknowledgment}

We would like to thank Troels Harmark, Aki Hashimoto, Charlotte Kristjansen, Niels Obers, Marta Orselli, Peter Ouyang, Costas Zoubos, and especially Ofer Aharony for helpful discussions. SH was supported by FNU via grant number 272-06-0434.
The work of OB was
supported in part by the Israel Science Foundation under grant
no.~568/05.

%%%%%%%%%%%%%%%%%%%%%%%%%%%%%%%%%%%%%
\appendix
\renewcommand{\theequation}{\Alph{section}.\arabic{equation}}
\setcounter{equation}{0}

\section{Alternative computation of the bulk Euler number}

The bulk contribution to the Euler number of $\mathbb{C}^4/\mathbb{Z}_k$
comes from the fixed point.
This can be computed in an alternative way using string theory by
compactifying on $T^8/\mathbb{Z}_k$. Of course
this is possible only for $k=2,3,4$ and $6$, since $\mathbb{Z}_k$ must be an 
automorphism of the lattice defining the torus $T^8$.
The Euler numbers of $T^8/\mathbb{Z}_k$ were computed in this way in \cite{Font:2004et},
and are shown in table (\ref{euler_numbers}).
We can then extract the Euler numbers of the corresponding fixed points by dividing
by the number of fixed points, taking care to take into account the fact that 
there are two and three types of fixed points respectively in the $\mathbb{Z}_4$ and $\mathbb{Z}_6$ cases.
\begin{table}[h]
\begin{center}
\begin{tabular}{|c|c|}
  \hline
   orbifold & $\chi$  \\
   \hline
  $T^8/\mathbb{Z}_2$ & 384 \\
  $T^8/\mathbb{Z}_3$ & 216 \\
  $T^8/\mathbb{Z}_4$ & 240 \\
  $T^8/\mathbb{Z}_6$ & 240 \\
  \hline
\end{tabular}
\caption{The Euler numbers of $T^8/\mathbb{Z}_k$ \cite{Font:2004et}.}
\label{euler_numbers}
\end{center}
\end{table}

For $k=2$ the compact orbifold has $2^8$ fixed points, therefore each one 
contributes an Euler number
\be
\label{Z2_euler_number}
\chi_{blk}(\mathbb{C}^4/\mathbb{Z}_2) = {384\over 2^8} = {3\over 2} = 2 - {1\over 2} \ .
\ee
For $k=3$ there are $3^4$ fixed points, and therefore
\be
\label{Z3_euler_number}
\chi_{blk}(\mathbb{C}^4/\mathbb{Z}_3) = {216\over 3^4} = {8\over 3} = 3 - {1\over 3}\ .
\ee

The compact $\mathbb{Z}_4$ orbifold has a total of $4^4$ fixed points.
Of those $2^4$ are fixed under $\mathbb{Z}_4$, and $4^4-2^4$ are fixed
under $\mathbb{Z}_2 \subset \mathbb{Z}_4$, and related pairwise to each other
under the generator of $\mathbb{Z}_4$.
The latter therefore correspond to $(4^4-2^4)/2$ $\mathbb{Z}_2$ fixed points.
Thus
\be 
\chi(T^8/\mathbb{Z}_4) = 240 = 2^4 \cdot \chi_{blk}(\mathbb{C}^4/\mathbb{Z}_4) 
+ {1\over 2}(4^4-2^4) \cdot \chi_{blk}(\mathbb{C}^4/\mathbb{Z}_2) \ ,
\ee
and using (\ref{Z2_euler_number}) we find
\be
\chi_{blk}(\mathbb{C}^4/\mathbb{Z}_4) = {15\over 4} = 4 - {1\over 4} \ .
\ee

The orbifold $T^8/\mathbb{Z}_6$ has three kinds of fixed points.
There is one point (the origin) fixed under the full $\mathbb{Z}_6$,
$3^4-1=80$ points fixed under the $\mathbb{Z}_3$ subgroup ($80/2$ doublets),
and $2^8-1=255$ points fixed under the $\mathbb{Z}_2$ subgroup
($255/3$ triplets). Therefore
\be 
\chi(T^8/\mathbb{Z}_6) = 240 = \chi_{blk}(\mathbb{C}^4/\mathbb{Z}_6) 
+ {80\over 2} \cdot \chi_{blk}(\mathbb{C}^4/\mathbb{Z}_3) 
+ {255\over 3}  \cdot \chi_{blk}(\mathbb{C}^4/\mathbb{Z}_2) \ ,
\ee
and using (\ref{Z2_euler_number}) and (\ref{Z3_euler_number}) we find
\be
\chi_{blk}(\mathbb{C}^4/\mathbb{Z}_6) = 6 - {1\over 6} \ .
\ee

%%%%%%%%%%%%%%%%%%%%%%%%%%%%%%%

\setcounter{equation}{0}
\section{ The parametrization of $S^7$ and $\mathbb{C}P^3$}
We define a 7-sphere $S^7$ by
\bea
X_1&=&\cos\xi\cos{\theta_1\over 2}e^{i{\psi_1+\phi_1\over 2}}\ ,\nn\\
X_2&=&\cos\xi\sin{\theta_1\over 2}e^{i{\psi_1-\phi_1\over 2}}\ ,\nn\\
X_3&=&\sin\xi\cos{\theta_2\over 2}e^{i{\psi_2+\phi_2\over 2}}\ ,\label{paramet}\\
X_4&=&\sin\xi\sin{\theta_2\over 2}e^{i{\psi_2-\phi_2\over 2}}\ ,\nn
\eea
where $0\le\xi\le\pi/2$, $0\le\psi_i<4\pi$, $0\le\phi_i<2\pi$, and $0\le\theta_i<\pi$, and 
$|X_1|^2+|X_2|^2+|X_3|^2+|X_4|^2=1$.
In terms of these coordinates, the $S^7$ metric takes the form 
\bea
ds_{S^7}^2&=&d\xi^2+{\cos^2\xi\over 4}\left\{(d\psi_1+\cos\theta_1d\phi_1)^2+d\theta_1^2+\sin^2\theta_1d\phi_1^2\right\}\nn\\
&&\hspace{0.7cm}+{\sin^2\xi\over 4}\left\{(d\psi_2+\cos\theta_2d\phi_2)^2+d\theta_2^2+\sin^2\theta_2d\phi_2^2\right\}\ .
\label{7spheremetric1}
\eea
We can further rewrite it as a $U(1)$ bundle over $\mathbb{C}P^3$
\be
ds_{S^7}^2=ds_{\mathbb{C}P^3}^2 + (d\varphi+\omega)^2\ ,
\ee
by introducing new coordinates
\be
\psi_1=2\varphi+\psi\ ,\qquad\qquad \psi_2=2\varphi-\psi\ .
\label{varphi}
\ee
The one-form takes the form
\be
\omega=\half(\cos^2\xi-\sin^2\xi)d\psi+\half\cos^2\xi\cos\theta_1d\phi_1+\half\sin^2\xi\cos\theta_2d\phi_2\ ,
\ee
and the $\mathbb{C}P^3$ metric is parametrized by
\bea
ds_{\mathbb{C}P^3}^2&=&d\xi^2+\cos^2\xi\sin^2\xi\left(d\psi+{\cos\theta_1\over 2}d\phi_1-{\cos\theta_2\over 2}d\phi_2\right)^2\nn\\
&&+{1\over 4}\cos^2\xi\left(d\theta_1^2+\sin^2\theta_1d\phi_1^2\right)
+{1\over 4}\sin^2\xi\left(d\theta_2^2+\sin^2\theta_2d\phi_2^2\right)\ .\label{CP3}
\eea
In this parametrization the $\mathbb{Z}_k$ orbifold is defined by the identifications
\be
\psi_i\sim\psi_i+4\pi/k\qquad\qquad(X_a\sim e^{2\pi i/k}X_a)\ ,
\label{orbifoldaction1}
\ee
which yields in terms of the new coordinates
\be
\varphi\sim\varphi+2\pi/k\ ,\qquad\qquad
\psi\sim\psi + 2\pi\ .
\label{orbifoldaction2}
\ee
The torsion 3-cycle, $S^3/\mathbb{Z}_k \subset S^7/\mathbb{Z}_k$, is defined by the $\xi=0$ surface
\be
|X_1|^2+|X_2|^2=1\ ,\quad X_3=X_4=0\ .\label{xi0}
\ee

\section{Conventions for RR fields and D-brane couplings}

The gauge invariant field strengths in (massless) Type IIA supergravity are given by:
\bea
H_3 &=& dB_2 \nonumber \\
F_2 &=& dC_1 \nonumber \\
\tilde{F}_4 &=& dC_3 - C_1\wedge H_3 \\
\tilde{F}_6 &=& -*\tilde{F}_4 = dC_5 - C_3\wedge H_3 \nonumber\\
\tilde{F}_8 &=& *F_2 = dC_7 - C_5\wedge H_3 \nonumber
\eea
Note that $**=-1$ for even forms in Minkowski spacetime, so $*\tilde{F}_6=\tilde{F}_4$
and $*\tilde{F}_8=-F_2$.
The Bianchi identities/equations of motion for the RR fields with sources are given by:
\bea
dF_2 &=&  - *j_7 \nonumber \\
d\tilde{F}_4 & = & *j_5 - F_2\wedge H_3 \nonumber \\
d*\tilde{F}_4 &=& -d\tilde{F}_6 = *j_3 + \tilde{F}_4\wedge H_3 \\
d*F_2 &=& d\tilde{F}_8 = *j_1 + *\tilde{F}_4\wedge H_3 \,, \nonumber
\eea
where we have used $H_3\wedge H_3 = 0$.
The sources for the RR fields come from the D-brane CS terms, which can be formally expressed as 
\cite{Green:1996dd,Cheung:1997az,Bachas:1999um} 
($\alpha' = 1$):
\be
S_{Dp}^{CS} = {1\over (2\pi)^p} \int_M \left[\sum_q C_q \wedge e ^{2\pi F + B_2} \wedge 
\sqrt{\hat{\cal A}(4\pi^2 R_N)\over \hat{\cal A}(4\pi^2 R_T)} \right]_{p+1}\,,
\ee
where it is understood that we keep all terms of total form degree $p+1$ on the RHS.
$\hat{\cal{A}}$ denotes the ``A-roof" (or Dirac) genus, which can be expressed in terms of
Pontrjagin classes:
\be
\hat{\cal A} = 1 - {1\over 24}p_1 + {1\over 5760}\left(7 p_1^2 -  4p_2\right) + \cdots \,,
\ee
where
\bea
p_1 &=& -{1\over 2(2\pi)^2} \Tr R^2 \\
p_2 & =& {1\over 8 (2\pi)^4} \left[(\Tr R^2)^2 - 2\Tr R^4 \right] \,,
\eea
and $R_T, R_N$ denote the curvatures
of the tangent and normal bundles to the D-brane worldvolume, respectively.
Note that our convention differs from that of \cite{Green:1996dd,Cheung:1997az,Bachas:1999um} by a relative sign
between the worldvolume gauge field contribution and the curvature contribution.
This is because in their convention the second Chern character ($ch_2$) for an ASD
connection is negative, whereas in our convention it is positive.

The contribution of a D$p$-brane to the source current associated with a D$q$-brane charge is therefore given 
in general by:
\be
*j_{q+1}^{(Dp)} = {\delta_{9-p}\over (2\pi)^{p-q}}\wedge \left[e^{2\pi F + B_2} 
\wedge \sqrt{\hat{\cal A}(4\pi^2 R_N)\over \hat{\cal A}(4\pi^2 R_T)} \right]_{p-q}\,,
\ee
where $\delta_{9-p}$ is a $\delta$-function localized at the spatial position of the D$p$-brane.
In particular for $q=2$,
\bea
*j_3^{(D4)} &=& {\delta_5\over (2\pi)^2} \wedge (2\pi F + B_2) \\
**j_3^{(D6)} &=& {\delta_3\over (2\pi)^4} \wedge \left[{1\over 2} (2\pi F + B_2)^2 
+ {(2\pi)^4\over 48}\left(p_1(TM) - p_1(NM)\right) \right] \,.
\eea

\end{document}